\documentclass[conference]{IEEEtran}

\IEEEoverridecommandlockouts
\usepackage{cite}
\usepackage{amsmath,amssymb,amsfonts}
\usepackage{algorithmic}
\usepackage{graphicx}
\usepackage{textcomp}
\usepackage{algorithm2e}
\usepackage{xcolor}
\def\BibTeX{{\rm B\kern-.05em{\sc i\kern-.025em b}\kern-.08em
    T\kern-.1667em\lower.7ex\hbox{E}\kern-.125emX}}

\usepackage{caption}
\usepackage{subcaption}
    
\usepackage{tikz}
\usetikzlibrary{quantikz}

\usepackage{subcaption}

\begin{document}

\title{Distributed Quantum Computing with Fan-Out Operations  and Qudits: the Case of  Distributed Global Gates}

\author{\IEEEauthorblockN{Seng W. Loke}
\IEEEauthorblockA{\textit{School of Information Technology, Deakin University, Burwood, VIC 3125, Australia.} \\
seng.loke@deakin.edu.au}
 }


\maketitle

\begin{abstract}
Much recent work on distributed quantum computing have focused on the use of entangled pairs  and distributed two qubit gates. But there has also been work on efficient schemes for achieving multipartite entanglement between nodes in a single shot, removing the need to generate multipartite entangled states  using many entangled pairs. 
This paper looks at how  multipartite entanglement resources (e.g., GHZ states) can be useful for distributed fan-out operations; we also consider the use of qudits of dimension four for distributed quantum circuit compression.
In particular, we consider how such fan-out operations and qudits can be used to implement  circuits which are challenging for distributed quantum computation, involving pairwise qubit interactions, i.e., what has been called global gates (a.k.a. global Mølmer-Sørensen gates). Such gates have been explored to possibly yield more efficient computations via reduced  circuit depth, and can be carried out efficiently in some types of quantum hardware (e.g., trapped-ion quantum computers); we consider this as an exploration of an ``extreme'' case for distribution given the global qubit-qubit interactions.  We also conclude with some implications for future work on quantum circuit compilation and quantum data centre design.
\end{abstract}


\section{Introduction}

The importance of multipartite entanglement for  quantum network applications, distributed large-scale quantum computers, and quantum data centres have been well recognized~\cite{Shapourian:2025ezs,Cacciapuoti:2025zfn,10.1145/3649329.3655908}. Indeed, there have been  experimental realizations of distributed two qubit gates (e.g., ~\cite{main25}\footnote{See also https://photonic.com/wp-content/uploads/2024/05/DQC-in-Si-May-Final.pdf [accessed: 17/10/25]}) but there have also been experimental realizations of distributed multipartite entanglement (e.g.,~\cite{Main:2025hda}); in particular, while multipartite entanglement can be achieved via many instances of bipartite entanglement (e.g., effectively fusing Bell-pairs or multipartite states) such as in~\cite{9292429,10313795}, there are also more efficient {\em one shot} approaches~\cite{Ainley:2024bdu}.  Such multipartite entanglement, e.g., GHZ states, are useful for distributed single control multitarget fan-out operations, and if such fan-out operations can be realized efficiently, assuming distributed multipartite entanglement for multiple nodes can be done efficiently in a one shot manner, then, distributed quantum circuits using such distributed fan-out operations (possibly with some dCNOTs and single qubit gates) can be more efficiently executed than equivalent distributed circuits using just distributed two-qubit operations (and entangled pairs) alone (with single qubit gates).


Global gates (a.k.a. global Mølmer-Sørensen gates~\cite{PhysRevLett.82.1971}, or GMS gates, for short) are natural and efficient for certain physical implementations such as trapped ion qubits where entire arrays of qubits can  be targeted and pairwise qubit-qubit interactions (i.e., global interactions between all or a subset of qubits) can be naturally realized, and have been explored for efficient quantum circuit constructions~\cite{maslovandnam2018}. Arbitrary circuits using single and two qubit gates can be transformed into an equivalent circuit with possibly reduced depth using such global gates and single-qubit gates as a gate set~\cite{Wetering2021,maslovandnam2018}. 

In this paper, we explore how to realize distributed versions of  global gates  using  distributed quantum circuits. We show how to implement distributed GMS  (in particular, GCZ) operations using GHZ states and qudit-based circuits, and explore implications for quantum data centre design.
The reason for exploring how to distribute GMS gates in this paper are: (i) distributed GMS gates have structure that map well to distributed fan-out operations which use GHZ states, (ii) given its pairwise interactions with qubits on different nodes, it would be a challenging operation to implement efficiently in a distributed manner - we wanted to see how far one could get in implementing such distributed operations, if we use fan-out operations and qudit-based circuit compression,  and  (iii) we wanted to consider if such distributed GMS gates (with single qubit gates) can form a gate set for efficient distributed quantum circuits with reduced depth; indeed, if using    distributed fan-out operations via one-shot multipartite entanglement and qudit-based circuit compression,  entanglement resources required for a distributed GMS gate can be less than that required for naive realizations using entangled pairs alone (assuming one-shot distributed GHZ states are not too much more resource-consuming  than  entangled pairs).

Section~\ref{sec1} first discusses using GHZ states for implementing distributed fan-out operations (single control multitarget operations).  Then,  Section~\ref{sec2} first introduces GMS gates, and then considers distributed versions of GMS gates (and a special instance, GCZ operations) and  their implementation using distributed fan-out operations with GHZ states. In Section~\ref{sec3} we consider qudit-based compression of qubit circuits such as for the GCZ operations, and consider generalizations of the distributed fan-out operations for qubits to  fan-out operations for qudits over two and three nodes, and show how such fan-out operations can   implement distributed GCZ operations. 
Section~\ref{sec4} concludes with future work.

\section{Distributed Fan-Out  Operations}
\label{sec1}
\begin{figure}[t]
  \includegraphics[width=0.48\textwidth]{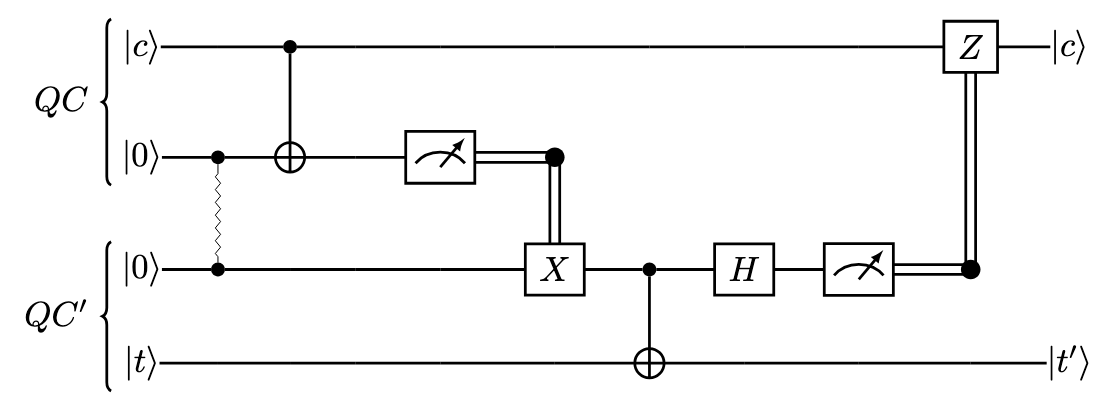}
\caption{Distributed control-$U$ (i.e., dCNOT is when U=$X$ ($\oplus$) as shown)   between nodes $QC$ and $QC'$ (the {\em computation} qubits are $c$, the control qubit, and $t$, the target qubit), and the wavy line illustrates a Bell pair  involving  the  {\em communication} qubits from the two nodes both initially $\ket{0}$.}
\label{dCNOT}
\end{figure}

In quantum circuits comprising only CNOT gates and single qubit gates, when such circuits are distributed, some of the CNOT gates are then converted into dCNOT gates~\cite{loke2023distributed, eisert}, and an entangled pair is needed as a resource for each dCNOT gate, as shown in Figure~\ref{dCNOT}.

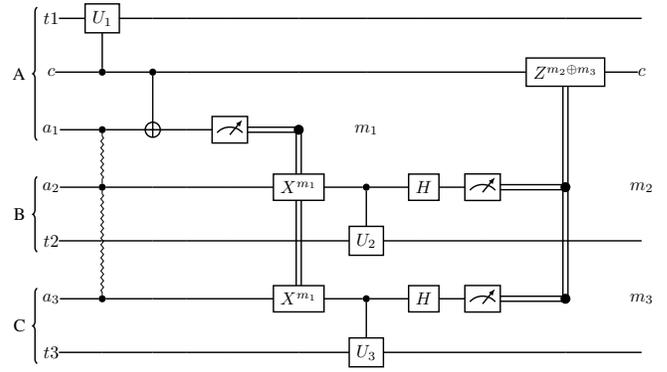
\begin{figure}

\begin{center}
\scalebox{.67}{
      \begin{quantikz} 
   \lstick[3]{A}          t1     \qw & \gate{U_1}             & \qw       & \qw    & \qw &  \qw &   \qw      & \qw           & \qw &  \qw &\qw       \\ 
   c                           \qw   & \ctrl{-1}  & \ctrl{1}  & \qw   & \qw & \qw & \qw &   \qw &  \qw   &   \gate{Z^{m_2\oplus m_3}} & \qw c  \\
    a_1    \qw & \ctrl{}{}  & \targ{}  & \qw & \meter{} &  \cwbend{3} & m_1         &            &  &         \\
    \lstick[2]{B}    a_2    \qw & \ctrl{}{}  & \qw       & \qw & \qw      & \gate{X^{m_1}}   & \ctrl{1} & \gate{H}& \meter{} &  \cwbend{-2}  &m_2  \\
      t2   \qw & \qw        & \qw       & \qw & \qw      & \qw         & \gate{U_2}       & \qw        & \qw &  \qw &\qw  \\
          \lstick[2]{C}    a_3   \qw & \ctrl{}{}  & \qw       & \qw & \qw      & \gate{X^{m_1}}   & \ctrl{1} & \gate{H}  & \meter{} &  \cwbend{-2}   & m_3      \\
      t3   \qw & \qw        & \qw       & \qw & \qw      & \qw         & \gate{U_3}       & \qw        & \qw &  \qw &\qw 
         \arrow[from=3-2,to=4-2,squiggly,dash,line width=0.1mm]{}
         \arrow[from=6-2,to=4-2,squiggly,dash,line width=0.1mm]{}
    \end{quantikz}
    }
\end{center}
\caption{Distributed multitarget control operation with single control qubit (on $A$) for multiple target qubits (one on $A$, one on $B$ and one on $C$)}
\label{multitarget}
\end{figure}

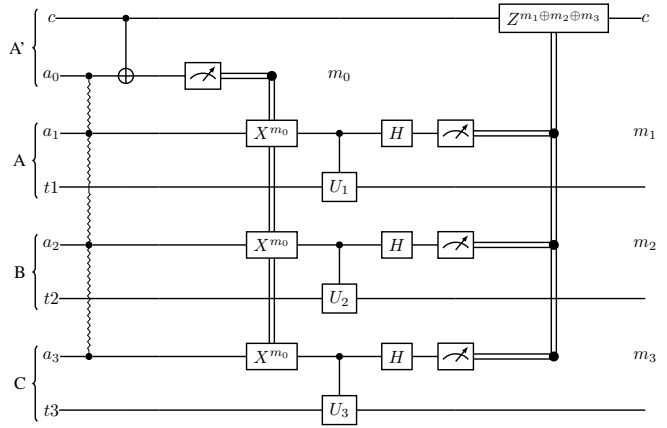
\begin{figure}

\begin{center}
\scalebox{.67}{
      \begin{quantikz} 
   \lstick[2]{A'}    c                           \qw   & \qw  & \ctrl{1}  & \qw   & \qw & \qw & \qw &   \qw &  \qw   &   \gate{Z^{m_1 \oplus m_2 \oplus m_3}} & \qw c  \\
    a_0    \qw & \ctrl{}{}  & \targ{}  & \qw & \meter{} &  \cwbend{5} &    m_0      &            &  &         \\
    \lstick[2]{A}    a_1    \qw & \ctrl{}{}  & \qw       & \qw & \qw      & \gate{X^{m_0}}   & \ctrl{1} & \gate{H}& \meter{} &  \cwbend{-2} &  m_1 \\
      t1   \qw & \qw        & \qw       & \qw & \qw      & \qw         & \gate{U_1}       & \qw        & \qw &  \qw &\qw   \\
          \lstick[2]{B}    a_2   \qw & \ctrl{}{}  & \qw       & \qw & \qw      & \gate{X^{m_0}}   & \ctrl{1} & \gate{H}  & \meter{} &  \cwbend{-2}   & m_2    \\
      t2   \qw & \qw        & \qw       & \qw & \qw      & \qw         & \gate{U_2}       & \qw        & \qw &  \qw &\qw  \\
                \lstick[2]{C}    a_3   \qw & \ctrl{}{}  & \qw       & \qw & \qw      & \gate{X^{m_0}}   & \ctrl{1} & \gate{H}  & \meter{} &  \cwbend{-2}   & m_3    \\
      t3   \qw & \qw        & \qw       & \qw & \qw      & \qw         & \gate{U_3}       & \qw        & \qw &  \qw &\qw  
         \arrow[from=2-2,to=4-2,squiggly,dash,line width=0.1mm]{}
         \arrow[from=7-2,to=4-2,squiggly,dash,line width=0.1mm]{}
    \end{quantikz}
    }
\end{center}
\caption{Distributed multitarget control operation with single control qubit (on $A'$) for multiple target qubits (one on $A$, one on $B$ and one on $C$) - all target qubits on different nodes from the control qubit.}
\label{multitarget2}
\end{figure}

We first consider generalized control operations where we point out that the use of multipartite states such as GHZ states can be potentially faster than using a sequence of dCNOT gates due to reduced circuit depth.
By a distributed fan-out CNOT operation, we refer to an operation where the same control qubit is used for multiple gate operations, on different nodes. This can arise from the structure of a quantum circuit itself, or from a control-$U$ operation where the unitary $U$ spans multiple qubits and is decomposed into operations executed on qubits distributed on diferent nodes~\cite{Yimsiriwattana:2004xhy,Neumann2020Imperfect,loke2023distributed,PhysRevA.107.L060601}. For example, the quantum circuit such as:
    \begin{center}
      \begin{quantikz}
     \ket{c}     & \ctrl{1} & \ctrl{2}  & \ctrl{3}  & \qw       \\
     \ket{t_1}     & \gate{U_1} &  \qw  & \qw  & \qw      \\
     \ket{t_2}     & \qw        &  \gate{U_2}   & \qw   & \qw    \\
     \ket{t_3}      & \qw     &  \qw  & \gate{U_3}  & \qw             \\
    \end{quantikz}
    \end{center}
i.e., a {\em multitarget control-$U$}, say  with decomposition $U=U_1 \cdot U_2 \cdot U_3$, where $U_1$ acts on qubit~$t1$, $U_2$ acts on qubit~$t2$, and $U_3$ acts on qubit~$t3$, each qubit on a different node, will result in the distributed CNOT as shown in Figure~\ref{multitarget}; note that the wavy lines in the figure connecting the three (communication) qubits $a_1,a_2,a_3$  represents a distributed GHZ state of the form $\frac{1}{\sqrt{2}}(\ket{0_A 0_B 0_C}+\ket{1_A1_B1_C})$, over three nodes $A$, $B$ and $C$. If the control and target qubits are all on different nodes, we can use a 4-qubit GHZ state, as shown in Figure~\ref{multitarget2} (qubits $a_0,a_1,a_2$ and $a_3$).

An alternative implementation not using a GHZ state, is to convert  the CNOTs into  dCNOTs, using entangled pairs, one for each dCNOT as shown in Figure~\ref{dCNOT}, but this can increase the circuit depth and time taken (especially  if the  dCNOTs are done sequentially, which might be required if the node $A'$ (with the control qubit) only has one communication qubit).
The advantage of the GHZ state approach could grow with the number of such CNOT gates (e.g., with $n$ such target qubits, on different nodes,  all different from the control qubit, we can use one ($n+1$)-qubit GHZ state instead of $n$  Bell pairs).

Suppose the time cost of one distributed n-qubit GHZ state is $t_{n-ghz}$, and  
the cost of establishing an entangled pair (between any pair of qubits) is $t_{ep}$ (this would be a gross simplification since entanglement swapping might be required for further apart pairs of qubits), and suppose the time cost for a single shot GHZ state $t_{n-ghz}$ = $\epsilon t_{ep}$ for some $\epsilon \approx 1$, then, for a multitarget control operation distributed with one qubit per node, we have time cost gain of
$n t_{ep} - t_{(n+1)-ghz} \approx  (n-1)t_{ep}$. That is, we have assumed $t_{n-ghz}$ similar with any $n$, so that we could be looking at almost constant time for using distributed GHZ states versus $O(n)$ resources using just dCNOT gates, though with larger $n$, we would expect $t_{n-ghz}$ to increase.


\section{Distributed GMS Gates}
\label{sec2}
\begin{figure}[h]
\centering
\includegraphics[width=1.0\linewidth]{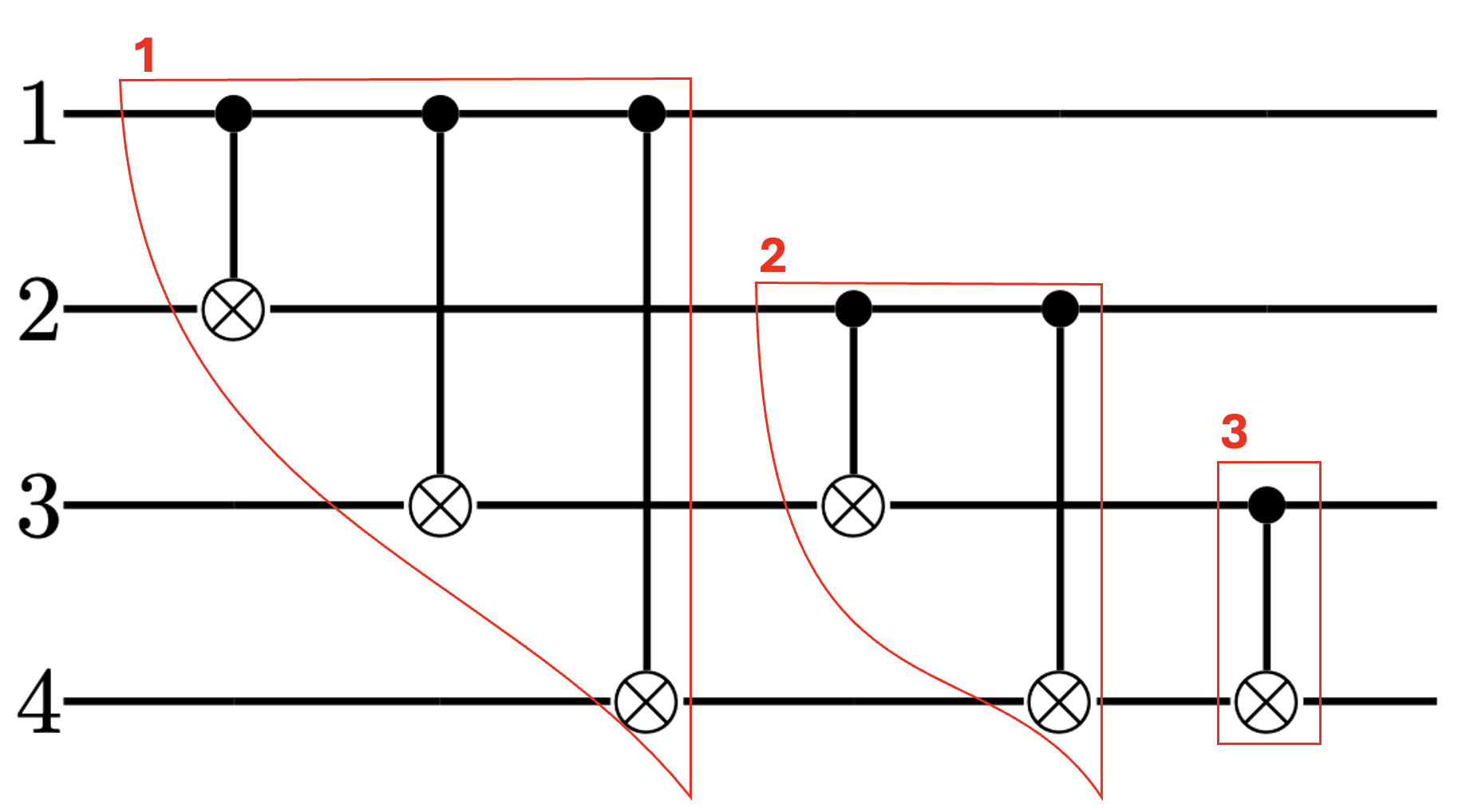}
\caption{A $GMS_{1,2,3,4}$ gate marked with triangles (fan-out~1 and fan-out~2) and box ({\em dLMS}~3) highlighting the qubits involved in a GHZ state for each triangle and a distributed control-U  between qubits 3 and 4, which would be needed if the qubits were   located on different nodes, one qubit per node.}
\label{GMSmarked}
\end{figure}

A global MS gate or operation (GMS) over a set of qubits $S$ is defined as follows: $GMS_S(\theta)=$
\[
 exp(-i\frac{\theta}{2} \sum_{i,j\in S, ~ i <j} X_i X_j)  = \prod_{i,j\in S, ~ i <j} exp(-i\frac{\theta}{2}X_i X_j)
 \]
where the operation $exp(-i\frac{\theta}{2}X_i X_j)$ is called the ``local'' MS gate acting on qubits $i$ and $j$ can be viewed operationally as: 
$exp(-i\frac{\theta}{2}X_i X_j)$ = \\
$(H_i \otimes H_j)  CNOT_{i\rightarrow j}  (I_i \otimes (R_Z(\theta))_j)  CNOT_{i\rightarrow j} (H_i \otimes H_j)$,
with  $R_Z(\theta) = \begin{bmatrix} e^{-i\frac{\theta}{2}} & 0 \\ 0 & e^{i\frac{\theta}{2}}  \end{bmatrix}$.
(Note that the local MS gate (or LMS gate, for short) is synmmetrical, i.e. $exp(-i\frac{\theta}{2}X_i X_j)$=$exp(-i\frac{\theta}{2}X_j X_i)$.)
For example, a $GMS_{1,2,3,4}$ gate involving four qubits can be visualized as shown in Figure~\ref{GMSmarked} (ignoring the triangles for now); in the figure, we use a symbol to represent a local MS gate, i.e.:
       \begin{center}
      \begin{quantikz}
          \ctrl{1}        \\
           \push{\otimes}   
    \end{quantikz}
    $\equiv$      \begin{quantikz}
    \gate{H} &   \ctrl{1} & \qw & \ctrl{1}        & \gate{H}  \\
    \gate{H} &   \targ{} & \gate{R_Z(\theta)} & \targ{} & \gate{H}   
    \end{quantikz}
    \end{center}

We consider how a distributed version  of the  above $GMS_{1,2,3,4}$ gate (called {\em dGMS} gate) might be implemented, considering one qubit per node setting (we leave out mentioning $\theta$ where convenient). 
One way to implement this is simply to use dCNOT gates instead of CNOT gates in each LMS gate (let's call a distributed version of such a LMS gate a {\em dLMS} gate), which will be two for each of the six {\em dLMS} gates in the $GMS_{1,2,3,4}$  gate yielding 12 dCNOT gates.

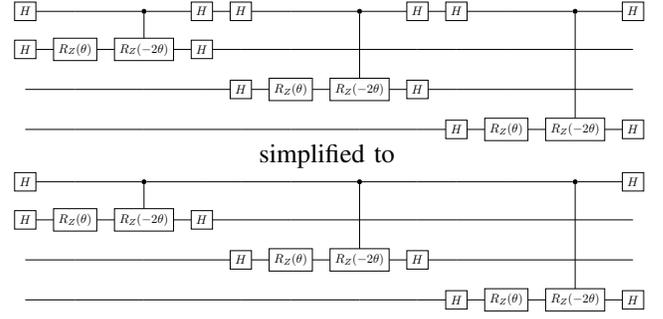
\begin{figure}
       \begin{center}
  \scalebox{0.47}{  \begin{quantikz}
    \gate{H} &    \qw & \ctrl{1}        & \gate{H} &   \gate{H} &   \qw & \ctrl{2}          & \gate{H} &   \gate{H} &   \qw  & \ctrl{3}        & \gate{H}   \\
    \gate{H} &   \gate{R_Z(\theta)} & \gate{R_Z(-2\theta)}   & \gate{H} & \qw  & \qw & \qw &\qw & \qw & \qw & \qw & \qw  \\
     \qw & \qw & \qw & \qw &   \gate{H} &   \gate{R_Z(\theta)} & \gate{R_Z(-2\theta)}   & \gate{H}  & \qw & \qw & \qw & \qw  \\
     & \qw & \qw & \qw  &\qw & \qw & \qw  & \qw  & \gate{H} &   \gate{R_Z(\theta)} & \gate{R_Z(-2\theta)}  & \gate{H} 
    \end{quantikz} } \\
  \mbox{simplified to}  \\
  \scalebox{0.47}{   \begin{quantikz}
    \gate{H} &    \qw & \ctrl{1}        & \qw &\qw &   \qw & \ctrl{2}          & \qw &\qw  &   \qw  & \ctrl{3}        & \gate{H}   \\
    \gate{H} &   \gate{R_Z(\theta)} & \gate{R_Z(-2\theta)}   & \gate{H} & \qw  & \qw & \qw &\qw & \qw & \qw & \qw & \qw  \\
     \qw & \qw & \qw & \qw &   \gate{H} &   \gate{R_Z(\theta)} & \gate{R_Z(-2\theta)}   & \gate{H}  & \qw & \qw & \qw & \qw  \\
     & \qw & \qw & \qw  &\qw & \qw & \qw  & \qw  & \gate{H} &   \gate{R_Z(\theta)} & \gate{R_Z(-2\theta)}  & \gate{H} 
    \end{quantikz} }
    \end{center}
\caption{The first three LMS operations in the $GMS_{1,2,3,4}$ gate showing that a fan-out operation can be used for the single qubit control for  $R_Z(-2\theta)$ in three different target qubits - the simplification is due to $HH=I$.}
\label{GMSfanout}
\end{figure}

However, suppose the hardware can do a conditional operation, denoted by $C(U_1,U_2)$, defined by:
\begin{align*}
C(U_1,U_2) & = \ket{0}\bra{0} \otimes U_1 + \ket{1}\bra{1} \otimes U_2 \\
& = (\ket{0}\bra{0} \otimes I + \ket{1}\bra{1} \otimes U_2U_1^{-1}) (I \otimes U_1)
\end{align*}
which  can be implemented using a single qubit control operation, and amenable to distributed control and fan-out operations as we have seen. Now, we observe that the LMS gate can be rewritten as: $exp(-i\frac{\theta}{2}X_i X_j)  =$
\begin{align*}
& (H \otimes H) \, (\ket{0}\bra{0} \otimes R_Z(\theta) + \ket{1}\bra{1} \otimes R_Z(-\theta)) \, (H \otimes H) \\
& = (H \otimes H)\,  C(R_Z(\theta),R_Z(-\theta)) \, (H \otimes H) 
\end{align*} 
or in circuit form: 
       \begin{center}
      \begin{quantikz}
          \ctrl{1}        \\
           \push{\otimes}   
    \end{quantikz}
    $\equiv$      \begin{quantikz}
    \gate{H} &   \qw & \ctrl{1}        & \gate{H}  \\
    \gate{H} &  \gate{R_Z(\theta)} & \gate{R_Z(-2\theta)}  & \gate{H}   
    \end{quantikz}
    \end{center}
noting that  $R_Z(-\theta)R_Z(\theta)^{-1} = R_Z(-\theta)R_Z(-\theta)=R_Z(-2\theta)$.
    So, observing now that the GMS gate has similarities with the fan-out operation we saw earlier, we could use fan-out operations, as illustrated in Figure~\ref{GMSfanout}  for the first three LMS gates of $GMS_{1,2,3,4}$.

So, we can use one 4-qubit GHZ state (involving qubits 1,2,3 and 4 for the first three distributed LMS gates) in a way similar to Figure~\ref{multitarget2}, one 3-qubit GHZ state (involving qubits 2, 3 and 4 for the next two LMS  gates) and one distributed control-U gate (involving qubits 3 and 4 for the last {\em dLMS} gate) for this (as illustrated in Figure~\ref{GMSmarked}), which could be more efficient if the one-shot method of establishing an $n$-qubit GHZ state (say, time $t_{n-ghz}$) takes only slightly longer than establishing an entangled pair (say $t_{ep}$), (without {\em entanglement caching}, i.e.,  storing a priori established entangled qubits), and can use  fewer communication qubits for parallelisation. We have: $2t_{n-ghz}+t_{ep} < 12t_{ep}$ (or $6t_{ep}$ if using the conditional operation approach), assuming $t_{n-ghz}$ = $\epsilon t_{ep}$ for some $\epsilon \approx 1$.  
In general, for a $GMS_S$ where $S$ is of size $n$, we would need $(n-2)$ GHZ states (involving $n$, $n-1$, ..., and $3$ qubits) for each multitarget control (or fan-out) operation, and one  entangled pairs (for the dLMS gate involving qubits  $n-1$ and $n$), or with just dCNOT gates, requiring $2n(n-1)/2=n(n-1)$ entangled pairs (or half that with the conditional operation approach), i.e. $O(n)$ resources using GHZ states instead of $O(n^2)$ resources with just entangled pairs.

When using this scheme, one can consider parallelising the fan-out operations, e.g., in the example in Figure~\ref{GMSmarked}, we try to perform fan-out~1 and fan-out~2 (and also the dLMS~3) concurrently. But note that the first LMS in fan-out~1 and  the first LMS in fan-out~2 may not commute, and so, just the first LMS in fan-out~1 must complete before the start of fan-out~2, which could be managed in node~2 (with qubit~2)  (one can envision a circuit where a 3-qubit GHZ state is used for fan-out~2 with qubit~2 as control, executed after its local LMS controlling on the communication qubit completes).  Hence, apart from the non-commuting first LMS gates of different fan-outs, some  concurrency is possible.

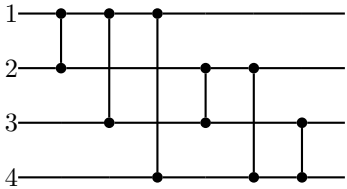
\begin{figure}[ht]
    \begin{center}
      \begin{quantikz}
     1     & \ctrl{1} & \ctrl{2}  & \ctrl{3}  & \qw  & \qw   & \qw & \qw   \\
     2     & \control{} &  \qw  & \qw  & \ctrl{1} & \ctrl{2}  & \qw & \qw     \\
     3     & \qw        &  \control{}   & \qw   & \control{}  & \qw  & \ctrl{1} & \qw   \\
     4      & \qw     &  \qw  & \control{}   & \qw  & \control{}          & \control{}  & \qw    \\
    \end{quantikz}
    \end{center}
    \caption{4-qubit $GCZ$ gate using pairwise CZ gates.}
\label{gcz}
\end{figure}

In the special cases where we are dealing with GCZ gates (equivalent up to Clifford gates to $GMS(\pi/2)$ gates~\cite{Wetering2021}\footnote{More precisely, $CZ  = e^{i\pi/4} (S^\dagger \otimes S^\dagger) (H \otimes H)\, e^{-i \frac{\pi}{4} X \otimes X} (H \otimes H)$, where $S^{\dagger} = e^{-i\pi/4}R_Z(-\frac{\pi}{2})$}), i.e., instead of a LMS gate, we apply a CZ gate between every pair of qubits, as illustrated in Figure~\ref{gcz}, the first CZ gate of the first fan-out (of CZ gates) then commutes with the first CZ gate of the second fan-out, etc., i.e., in a distributed version of a 4-qubit GCZ over four nodes, one qubit per node, we can perform the two fan-outs (using one 4-qubit distributed GHZ state involving qubits 1, 2, 3 and 4, and one 3-qubit GHZ state involving qubits 2,3 and 4) and the final LMS gate (using a distributed CZ, or dCZ gate) concurrently (as long as the GHZ states are available in parallel with enough communication qubits to support this). The circuit for the dCZ and these fan-outs  are similar to those in Figures~\ref{dCNOT} and~\ref{multitarget2}, but using the equivalence: $CZ = (I \otimes H) CNOT (I \otimes H)$ (i.e., adding $H$ gates on target nodes, or effectively use $CZ$ on target qubits).

\begin{figure}[ht]
  \begin{center}
    \begin{subfigure}[t]{0.45\textwidth}
        \includegraphics[width=\textwidth]{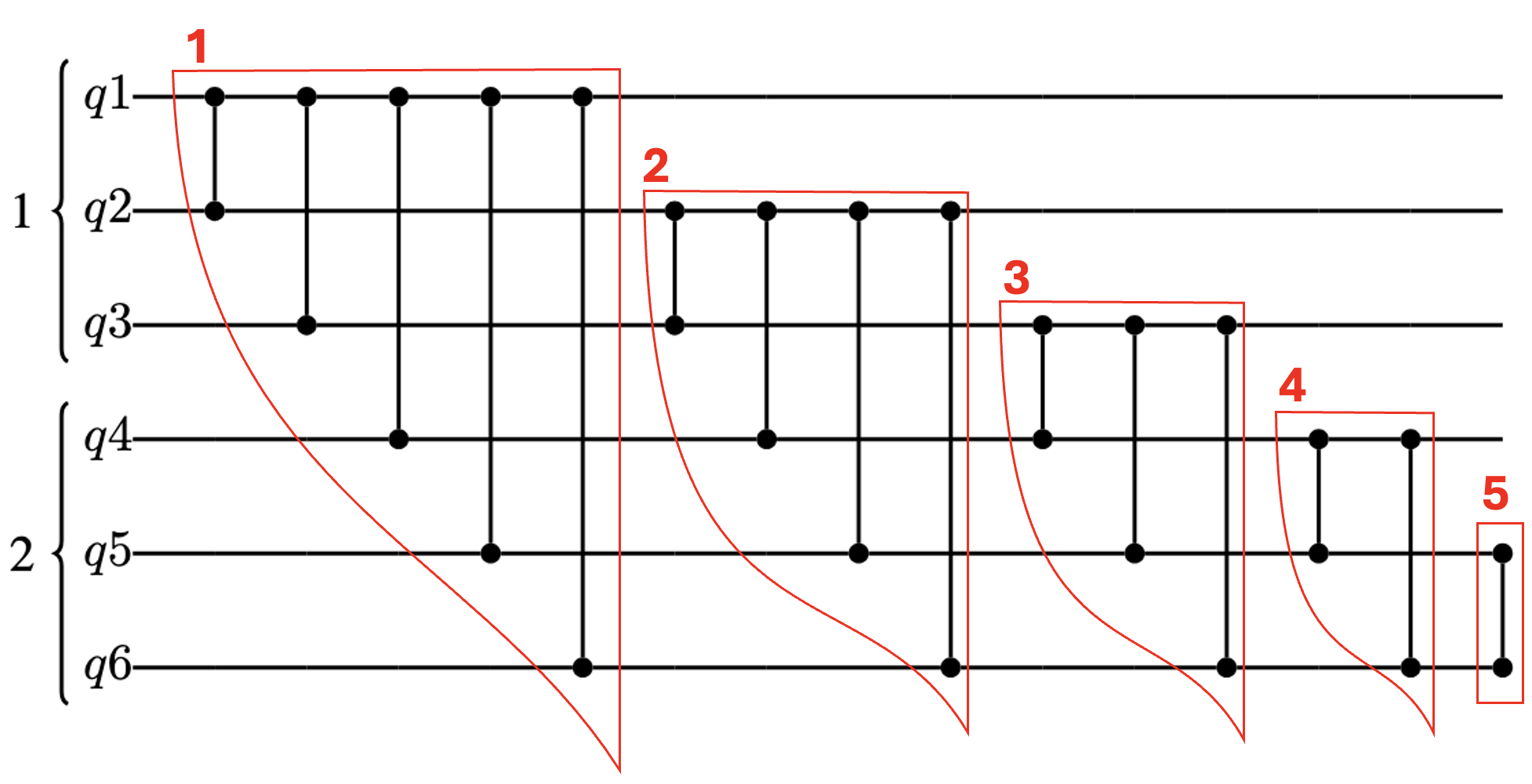}
        \caption{Distribution over two nodes.}
        \label{gcz-d3}
    \end{subfigure}
  \begin{subfigure}[t]{0.45\textwidth}
        \includegraphics[width=\textwidth]{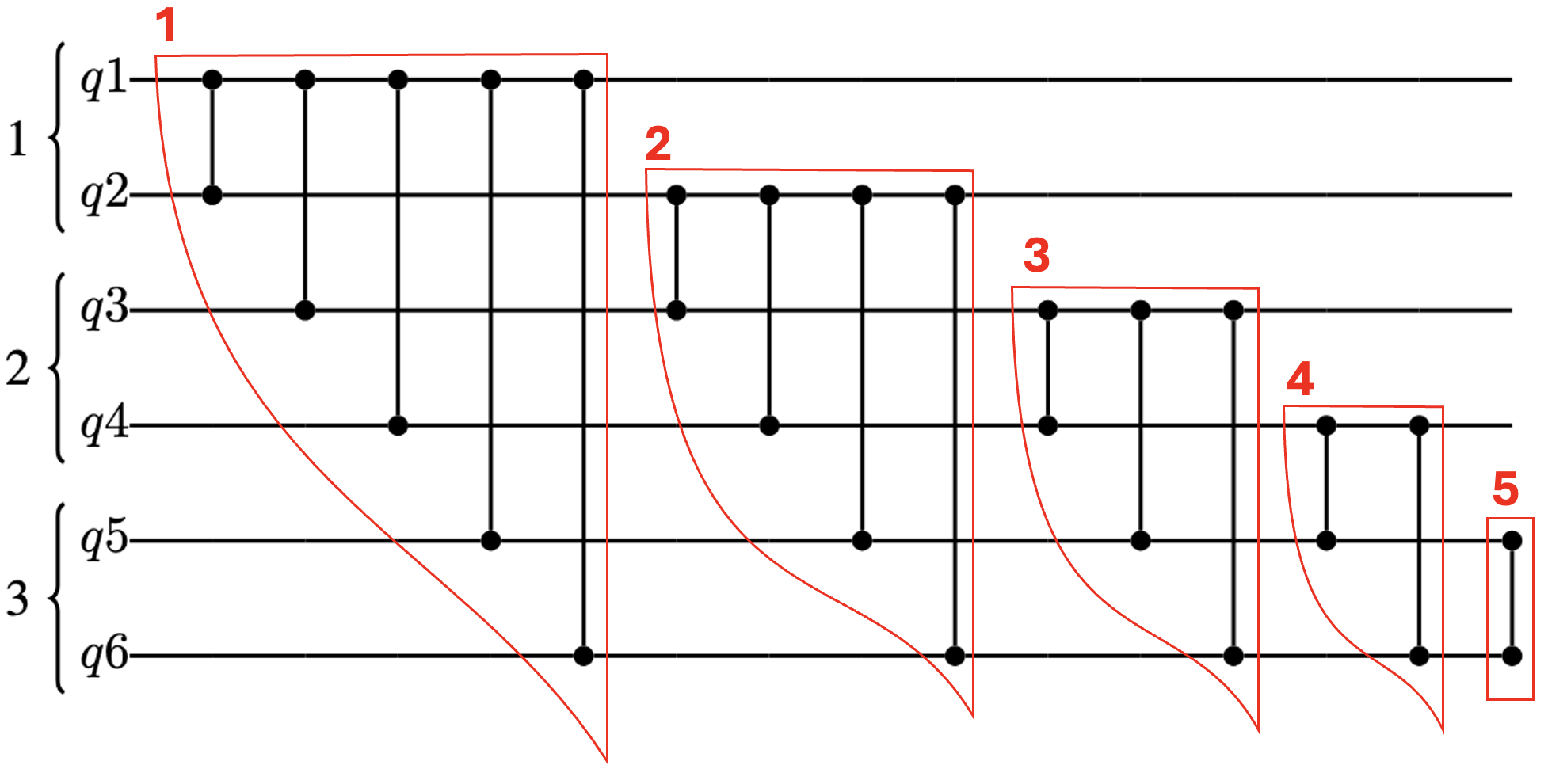}
        \caption{Distribution over three nodes.}
        \label{gcz-d2}
    \end{subfigure}
 \end{center}
    \caption{Implementation of a six qubit $GCZ$ gate using pairwise CZ gates, with a distribution over two and three nodes. }
\label{gcz6}
\end{figure}

So far we have only considered one qubit per node, we now consider multiple qubits allocated to each node, for a 6-qubit $GCZ$ gate as shown in Figure~\ref{gcz6}, over two and three nodes.

\begin{figure}
\begin{center}
\scalebox{.62}{
      \begin{quantikz}
         \lstick[4]{1}   q_2  & \targ{}  & \qw   & \qw  & \qw   & \qw & \qw & \qw &          \qw &  \qw   &   \qw & \qw &   \qw & \qw  \ket{q_2'}\\
    q_3 &  \qw   & \targ{}   & \qw  & \qw   & \qw & \qw & \qw &          \qw &  \qw   &   \qw & \qw  &   \qw & \qw  \ket{q_3'}\\
        q_1 &  \ctrl{-2}   & \ctrl{-1}   & \ctrl{1}  & \qw   & \qw & \qw & \qw &          \qw &  \qw  & \qw  &   \qw &   \gate{Z}  & \qw \ket{q_1} \\
    \ket{0}  &   \qw & \ctrl{}{}  & \targ{}   & \qw & \meter{} &  \cwbend{1} &          &            &  &   &   &        \\
    \lstick[4]{2}    \ket{0} &   \qw & \ctrl{}{}  & \qw       & \qw & \qw      & \gate{X}    & \ctrl{1} & \ctrl{2} & \ctrl{3}  & \gate{H} & \meter{} &  \cwbend{-2}  &     \\
       q_4  & \qw & \qw        & \qw       & \qw & \qw      & \qw         & \targ{}       & \qw        & \qw &  \qw &\qw   &   \qw & \qw \ket{q_4'} \\
           q_5  & \qw & \qw        & \qw       & \qw & \qw      & \qw         &  \qw      & \targ{}        & \qw &  \qw &\qw  &   \qw & \qw  \ket{q_5'}\\
        q_6  & \qw & \qw        & \qw       & \qw & \qw      & \qw         &   \qw    & \qw        & \targ{} &  \qw &\qw   &   \qw & \qw \ket{q_6'}
         \arrow[from=5-3,to=4-3,squiggly,dash,line width=0.1mm]{}
    \end{quantikz}
    }
\end{center}
    \caption{Circuit for implementing triangle~1 in Figure~\ref{gcz-d3} with control qubit $q_1$ - adding $H$ gates before and after the CNOT operations (on target qubits) will result in CZ operations, or effectively use $CZ$ on target qubits.}
\label{dCNOT-gcd-d3}
\end{figure}
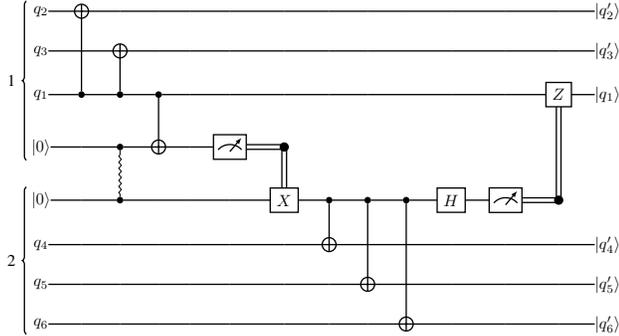

Over two nodes (Figure~\ref{gcz-d3}), for each distributed fan-out operation (triangles marked 1, 2, and 3), we can use a circuit similar to Figure~\ref{dCNOT} when there are some target qubits   on the other node,  where some of the target qubits are locally CNOT connected to the control qubit, and the target qubits on the other node are  linked to the control qubit via an entangled pair (e.g.,  Figure~\ref{dCNOT-gcd-d3} shows the circuit for triangle~1); using fan-out, in all, 3 entangled pairs of qubits are required, compared to using 9 entangled pairs if each CZ gate between qubits on different nodes were implemented separately (though another way in this case is to teleport all three qubits from one node to the other, perform all operations locally, and teleport the three qubits back, requiring 6 entangled pairs, one for each teleportation operation, but this requires a node with enough capacity to support computing with the larger number of qubits). 
Fan-out operation 4 (triangle 4) and the last CZ operation (box~5) are done locally on node~2 - in fact, both these operations together form a 3-qubit GCZ operation which can be done in one step on hardware that supports this.

Over three nodes (Figure~\ref{gcz-d2}),  2 GHZ states are required, both involving three nodes (for triangles 1 and 2 with controls $q_1$ and $q_2$ on node~1, respectively), each such triangle using a circuit similar to Figure~\ref{multitarget} (but extended to control two target qubits on each node), and  2 entangled pairs for operations marked with triangles~3 and 4, involving two nodes.  Provided the forming of the two GHZ states can be done efficiently (or similar time as an entangled pair), then this is favourable compared to 12 entangled pairs required if each  CZ gate between qubits on different nodes were implemented separately.
Another way, instead of using 12 entangled pairs, would be to teleport all qubits to one node and perform all operations locally and teleport the qubits back (requiring 4x2=8 teleportation operations), but again requiring a higher capacity node to accommodate all qubits.
Three  CZ operations are done locally on each node (between qubits $q_1$ and $q_2$ on node~1, $q_3$ and $q_4$ on node~2 and $q_5$ and $q_6$ on node~3).

  Next, we consider using qudits for  compressing the circuit for GCZ, and assuming the ability to compute with qudits and transfer qudit states, we discuss its potential.

\section{Qudit-Based Compression of Qubit Circuits and Distributed Qudit-Based Computing}
\label{sec3}
If the 4-qubit GCZ circuit in Figure~\ref{gcz} is distributed across two nodes (instead of four nodes we saw above), with the first two qubits on one node and the second two on another node, then, the implementation can use four distributed CZ gates (or dCZ, for short), each of which can be implemented using the circuit for the dCNOT gate with additional Hadamard gates to realize a dCZ (i.e., using $(I \otimes H)CNOT_{ij}(I \otimes H)=CZ_{ij}$). We consider what happens if we encode the first two qubits as a 4-dimensional qudit and the second two as another such qudit, For four qubits 1, 2, 3 and 4, encoded into two qudits, we can use the mapping:
$\ket{q_1q_2} \rightarrow  \ket{2q_1+q_2}$
and $\ket{q_3q_4} \rightarrow  \ket{2q_3+q_4}$.
Now, note that the GCZ circuit in Figure~\ref{gcz} computes the following:
\[
\sum_{q_i \in \{0,1\}} \alpha_{q_1q_2q_3q_4} (-1)^{p} \ket{q_1q_2}\ket{q_3q_4} 
\]
where $p=q_1q_2\oplus q_1q_3\oplus q_1q_4 \oplus q_2q_3 \oplus q_2q_4 \oplus q_3q_4$, (where $\oplus$ is addition mod~2),
which can be
rewritten as $p=q_1q_2 \oplus (q_1\oplus q_2)(q_3\oplus q_4) \oplus q_3q_4$.
This implies two intra-qudit operations 
and an inter-qudit operation. 

So, with the qudit encoding, we can use qudit operations (e.g., as mentioned in~\cite{Pudda:2024nrw,10.3389/fphy.2020.589504,Lysaght:2024omz,e26121129}) to implement an equivalent of the $GCZ$, including the following operations:
\begin{enumerate}
\item \[
P_3(\pi) = (I-\ket{3}\bra{3})+ (-1)\ket{3}\bra{3} 
\]
 which adds a phase $e^{i\pi} = -1$ to only  level $\ket{3}$, leaving other levels the same, useful for realizing an intra-qudit CZ operation adding (-1) only when $q_i q_j = 11$;
\item \[
X_{23} = \ket{0}\bra{0}+\ket{1}\bra{1}+\ket{2}\bra{3} + \ket{3}\bra{2}
\]
which is one of many possible $X_{ij}$ gates that generalizes the $X$ gate for a qubit, inter-changing two qudit states $\ket{i}$ and $\ket{j}$, leaving other states the same,\footnote{In this case, since the binary representation of 2 is 10 and 3 is 11, changing 10 to 11, and 11 to 10 is effectively a CNOT between two qubits being represented, with   00 and 01  unaffected, i.e., the computation gives $\ket{q_i}\ket{q_j} \rightarrow \ket{q_i}\ket{q_i \oplus q_j}$, giving the parity  $q_i \oplus q_j$.} and so if we apply $X_{23}$ on both qudits, we get the parities for $q_1 \oplus q_2$ and $q_3 \oplus q_4$ in $q2$ and $q4$, respectively;
\item  a generalization of the CZ gate:
\[
CZ_4 = \sum_{j,k\in \{0,1,2,3 \}} \omega^{jk}\ket{j}\ket{k} \bra{j}\bra{k}
\]
where $\omega = e^{{2\pi i}/d}= e^{{2\pi i}/4}= i$, and $jk = jk~(mod~4)$ where if we square this, we get:
\begin{align*}
(CZ_4)^2 & = \sum_{j,k\in \{0,1,2,3 \}} (-1)^{jk}\ket{j}\ket{k} \bra{j}\bra{k}
\end{align*}
\end{enumerate}

Now, note that:
{\small 
\begin{align*}
& (CZ_4)^2 ~ (X_{23} \otimes X_{23} ) \sum_{q_i \in \{0,1\}} \alpha_{q_1q_2q_3q_4} \ket{2q_1 + q_2}\ket{2q_3 + q_4} \\
& =  (CZ_4)^2 \sum_{q_i \in \{0,1\}} \alpha_{q_1q_2q_3q_4} \ket{2q_1 + (q_1 \oplus q_2)}\ket{2q_3 + (q_3\oplus q_4)} \\
& = \sum_{q_i \in \{0,1\}} \alpha_{q_1q_2q_3q_4} (-1)^{r}   \ket{2q_1 + (q_1 \oplus q_2)}\ket{2q_3 + (q_3\oplus q_4)}\\
&  \mbox{(where $r = (2q_1 + (q_1 \oplus q_2))(2q_3 + (q_3\oplus q_4))$)} \\
& = \sum_{q_i \in \{0,1\}} \alpha_{q_1q_2q_3q_4} (-1)^{r'}  \ket{2q_1 + (q_1 \oplus q_2)}\ket{2q_3 + (q_3\oplus q_4)}\\
&  \mbox{(where $r'  = (q_1\oplus q_2)(q_3\oplus q_4)$)} \\
\end{align*}
}
where the last equation is due to raising -1 to even powers.
Putting the above together, by applying the operation
\[
(P_3(\pi) \otimes P_3(\pi)) ~ (X_{23} \otimes X_{23} )~(CZ_4)^2 ~ (X_{23} \otimes X_{23} )
\]
to the qudits, we effectively compute:
\[
 \sum_{q_i \in \{0,1\}} \alpha_{q_1q_2q_3q_4} (-1)^{p}  \ket{2q_1 + q_2}\ket{2q_3 + q_4}\\
\]
which is what we expect of the GCZ gate. 

\subsection{Distributed Two Qudit Controlled Gates over Two Nodes and Distributed GCZ}
Note that for the two qudits on two different nodes, the intra-qudit operations are local, but the distributed version of the inter-qudit operation $(CZ_4)^2$ needs to be done, say  using a shared entangled pair of qudits,  with a circuit similar to Figure~\ref{dCNOT}, but measuring outcome $m \in \{0,1,2,3\}$ then informs an operation $X_4 \ket{j} = \ket{j+1~(mod~4)}$,
a $Z^{\dagger}_4$ operation: $Z^{\dagger}_4 \ket{j}= \omega^{-j}\ket{j}$, and instead of CNOT, using 
$CSUM_4: \ket{i} \ket{j}  \rightarrow \ket{i}\ket{(i+j) ~mod~4}$,  its inverse:
$CSUM_4^{\dagger}: \ket{i} \ket{j}  \rightarrow \ket{i}\ket{(j-i) ~mod~4}$,
and an operator to provide the modulo $d=4$ complement of the input: $K_4 \ket{i} = \ket{(4-i)~mod~4}=\ket{-i}$,
and instead of $H$ gate, use the operator:
$H_4 \ket{j} = \frac{1}{2} \sum_{k=0}^{3} \omega^{jk} \ket{k}$, and its inverse: $H^{\dagger}_4 \ket{j} = \frac{1}{2} \sum_{k=0}^{3} \omega^{-jk} \ket{k}$.

Let's first consider a distributed version of the $CSUM_4$ operation, which call $dCSUM_4$, involving two qudits on two different nodes $Q_1 = \sum_{j=0}^3 \alpha_j \ket{j}$ and $Q_2= \sum_{l=0}^3 \beta_l \ket{l}$ on nodes 1 and 2 respectively. 

\begin{figure}
\begin{center}
\scalebox{.55}{
      \begin{quantikz}
   \lstick[2]{1}   Q_1   \qw   & \qw\slice{1}   & \gate[2]{CSUM_4{^\dagger}} &\qw  & \qw   & \qw & \qw & \qw &          \qw &  \qw   &   \gate{(Z^{\dagger}_4)^{m'}} &  \qw  \ket{Q_1} \\
    E_1     \qw & \ctrl{}{}  & \slice{2}  & \gate{K_4} \slice{2'} & \qw & \meter{} &  \cwbend{1} & m &          &            &     &      \\
    \lstick[2]{2}    E_2    \qw & \ctrl{}{}  & \qw  &\qw      & \qw & \qw      & \gate{X_4^m}\slice{3}    & \gate[2]{CSUM_4}\slice{4} & \gate{H_4}\slice{5} & \meter{} &  \cwbend{-2}  & m'     \\
       Q_2   \qw & \qw        & \qw     &\qw   & \qw & \qw      & \qw         & \targ{}       & \qw        & \qw &  \qw &\qw \ket{Q_2'} 
         \arrow[from=3-2,to=2-2,squiggly,dash,line width=0.1mm]{}
    \end{quantikz}
    }
\end{center}
\caption{$dCSUM_4$ between $Q_1$ and $Q_2$ using an entangled pair of qudits $E_1 E_2$, with a possible update to qubit $Q_2$}
\label{dCSUM4}
\end{figure}
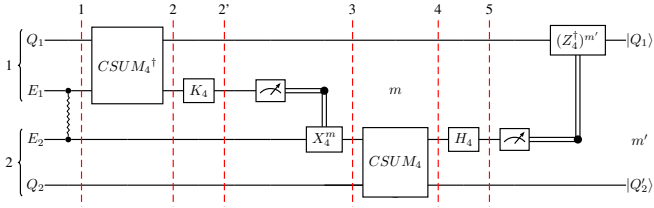

There has been previous work on a distributed generalized CNOT for qudits~\cite{chen07}, but we present our approach here analogous to Figure~\ref{dCNOT}; the circuit is illustrated in Figure~\ref{dCSUM4}.
Assume an entangled pair of qudits of the form $\sum_{k\in \{0,1,2,3\}} \frac{1}{2}\ket{kk}$ shared between the two nodes involving qudits $E_1$ and $E_2$ on nodes 1 and 2 respectively (at slice~1):
\[
Q_1 \otimes \sum_{k\in \{0,1,2,3\}} \frac{1}{2}\ket{kk}_{E_1 E_2} \otimes Q_2
\]

After a local $(CSUM_4^{\dagger})_{Q_1 \rightarrow E_1}$ between $Q_1$ and $E_1$, we obtain:
\[
\frac{1}{2}  \sum_{j,k \in \{0,1,2,3\}} \alpha_j \ket{j}_{Q_1}\ket{(k-j)~mod~4}_{E_1}\ket{k}_{E_2} \otimes Q_2
\]
(slice~2), and applying $K_4$ to $E_1$, we get
\[
\frac{1}{2} \sum_{j,k \in \{0,1,2,3\}} \alpha_j \ket{j}_{Q_1}\ket{(j-k)~mod~4}_{E_1}\ket{k}_{E_2} \otimes Q_2
\]
(slice~2'). Then, measuring $E_1$, we get outcome $m \in \{0,1,2,3\}$ (where the states $Q_1$ and $E_2$ are related by  $m=(j-k)~ mod~4$), encoded with 2 bits which are then sent to the other node (node~2).

On receiving $m$, the node applies 
$X_4^{m}$  to $E_2$ to get:
\[
\sum_{m=j-k} \alpha_j \ket{j}_{Q_1}\ket{(k+m)~mod~4}_{E_2} \otimes Q_2
\]
omitting the measured  $E_1$ (slice~3), but since $m= (j-k)~mod~4$, and given $Q_2$ above, we actually have:
\[
\sum_{j \in \{0,1,2,3\}} \alpha_j \ket{j}_{Q_1}\ket{j}_{E_2} \otimes (\sum_{l=0}^3 \beta_l \ket{l})_{Q_2}
\]

Now apply a $(CSUM_4)_{E_2 \rightarrow Q_2}$ to get:
\[
\sum_{j,l \in \{0,1,2,3\}} \alpha_j \beta_l \ket{j}_{Q_1}\ket{j}_{E_2} \ket{(j+l)~mod~4}_{Q_2}
\]
(slice~4). Then, we apply $H_4$ on $E_2$ to get:
\[
\sum_{j,l} \alpha_j \beta_l \ket{j}_{Q_1} (\frac{1}{2}\sum_{r=0}^3 \omega^{jr}\ket{r})_{E_2} \ket{(j+l)~mod~4}_{Q_2}
\]
(at slice~5). We then measure $E_2$ to get $m'$ (2 bits) which are then sent back to node~1, i.e. we have:
\[
\sum_{j,l} \omega^{jm'} \alpha_j \beta_l \ket{j}_{Q_1}  \ket{(j+l)~mod~4}_{Q_2}
\]
(omitting the measured $E_2$). Node~1 then applies $(Z^{\dagger}_4)^{m'}$ on $Q_1$ to get:
\begin{align*}
& \sum_{j,l} \omega^{-jm'} \omega^{jm'} \alpha_j \beta_l \ket{j}_{Q_1}  \ket{(j+l)~mod~4}_{Q_2} \\
& = \sum_{j,l}  \alpha_j \beta_l \ket{j}_{Q_1} \ket{(j+l)~mod~4}_{Q_2}
\end{align*}
as required.

Now we can use the above $dCSUM_4$ to implement a distributed $CZ_4$ (or $dCZ_4$) since $CZ_4 = (I \otimes H_4)~CSUM_4~(I \otimes H_4^{\dagger})$.
by adding gates $H_4$ and $H_4^{\dagger}$ just before and just after the $CSUM_4$ in $Q_1$ at node~2. Then, to realize a distributed $(CZ_4)^2$, (call this $d(CZ_4)^2$  operations  for short) we can repeat the $CZ_4$ circuit, i.e.,  at node~2, in Figure~\ref{dCSUM4}, we replace the $CSUM_4$ with $(CZ_4)^2$, or use the simplified $(I \otimes H_4)~CSUM_4~(I \otimes H_4^{\dagger})(I \otimes H_4)~CSUM_4~(I \otimes H_4^{\dagger})$ = $(I \otimes H_4)~CSUM_4~CSUM_4~(I \otimes H_4^{\dagger})$.

Hence,  the 4-qubit $GCZ$ gate would require four dCNOT gates (or four entangled pairs of qubits) when qubits 1 and 2 are on one node and qubits 3 and 4 are on another node, but using  two qudits, one on each node, we can compute an equivalent of the $GCZ$ using one entangled pair of qudits - if the time/cost resources for forming one entangled pair of qudits is similar to that of forming one entangled pair of qubits, this would be a significant reduction in resources. With higher dimensional qudits, there could be greater reduction in resources required. 

\subsection{Distributed Qudit-Based Fan-Out Operations over Three Nodes and Distributed GCZ}
We can extend this idea to distribution of $GCZ$ gates over three or more nodes,  using qudit versions of  $GHZ$ states. For example, for a   6-qubit $GCZ$ gate distributed over three nodes (e.g., 2 qubits per node), one can construct a circuit  analogous to Figure~\ref{multitarget}, with each pair of qubits on each node compressed into a 4-dimensional qudit, and the corresponding use of a 3-qudit GHZ state and an entangled pair of qudits.

Note that a 6-qubit GCZ circuit computes the following:
\[
\sum_{q_i \in \{0,1\}} \alpha_{q_1q_2q_3q_4q_5q_6} (-1)^{p} \ket{q_1q_2}\ket{q_3q_4} \ket{q_5q_6}
\]
where $p=q_1q_2\oplus q_1q_3\oplus q_1q_4 \oplus q_1q_5\oplus q_1q_6  \oplus q_2q_3 \oplus q_2q_4 \oplus q_2q_5 \oplus q_2q_6 \oplus q_3q_4  \oplus q_3q_5  \oplus q_3q_6  \oplus q_4q_5 \oplus q_4q_6 \oplus  q_5q_6$, which could be rewritten as
$p=q_1q_2\oplus (q_1\oplus q_2)(q_3\oplus q_4)   \oplus (q_1\oplus q_2)(q_5 \oplus q_6) \oplus q_3q_4  \oplus (q_3 \oplus q_4)(q_5\oplus q_6)   \oplus  q_5q_6$.  With an encoding of the first two qubits in one qudit, the second two in another qudit and the third pair in a third qudit, and each pair of qubit (or each qudit) on a different node (a total of three nodes),  this suggests the intra-qudit operations related to $q_1q_2$, $q_3q_4$ and  $q_5q_6$, and inter-qudit operations between nodes 
1 and 2 in relation to $(q_1\oplus q_2)(q_3\oplus q_4)$, and 1 and 3 in relation to $(q_1\oplus q_2)(q_5 \oplus q_6)$, and between nodes 2 and 3 for $(q_3 \oplus q_4)(q_5\oplus q_6)$.  We can use three entangled pairs of qudits, one for each inter-qudit operation.

\begin{figure}
\begin{center}
\scalebox{.52}{
      \begin{quantikz} 
   \lstick[2]{1}    Q_1                           \qw   & \qw  & \gate[2]{CSUM_4{^\dagger}} & \qw   & \qw & \qw & \qw &   \qw &  \qw   &   \gate{(Z^{\dagger}_4)^{m_1 + m_2}} & \qw \ket{Q_1}  \\
    E_1    \qw & \ctrl{}{}  &   &\gate{K_4}  & \meter{} &  \cwbend{3} &    m_0      &            &  &         \\
    \lstick[2]{2}    E_2    \qw & \ctrl{}{}  & \qw       & \qw & \qw      & \gate{X_4^{m_0}}   & \gate[2]{CSUM_4} & \gate{H_4}& \meter{} &  \cwbend{-2} &  m_1 \\
      Q_2   \qw & \qw        & \qw       & \qw & \qw      & \qw         &        & \qw        & \qw &  \qw &\qw  \ket{Q_2'} \\
          \lstick[2]{3}    E_3   \qw & \ctrl{}{}  & \qw       & \qw & \qw      & \gate{X_4^{m_0}}   & \gate[2]{CSUM_4} & \gate{H_4}  & \meter{} &  \cwbend{-2}   & m_2    \\
      Q_3   \qw & \qw        & \qw       & \qw & \qw      & \qw         &        & \qw        & \qw &  \qw &\qw  \ket{Q_3'}
         \arrow[from=2-2,to=4-2,squiggly,dash,line width=0.1mm]{}
         \arrow[from=5-2,to=4-2,squiggly,dash,line width=0.1mm]{}
    \end{quantikz}
    }
\end{center}
\caption{Distributed single control multitarget $CSUM''_4$ operation with single control qudit $Q_1$ (on node~1) for two target qudits ($Q_2$ on node~2 and $Q_3$ on node~3).}
\label{multitarget2-dcsum4}
\end{figure}
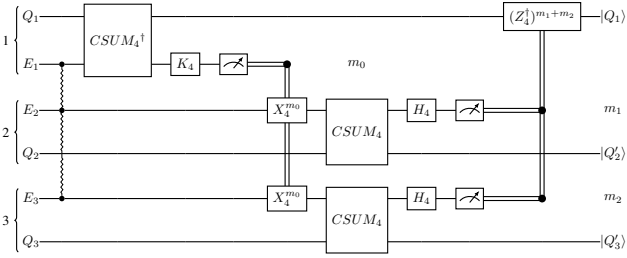

However, since node 1 needs to communicate  its parity  $(q_1\oplus q_2)$ to both nodes 2 and 3, we can construct a single control multiple target variation of $CSUM_4$, analogous to our single control multitarget CNOT; we call it $CSUM''_4$, i.e.:
\[
CSUM''_4: \ket{i} \ket{j} \ket{k} \rightarrow \ket{i} \ket{(i+j) mod~4} \ket{(i+k) mod~4}
\]
This would then enable us to implement the required  $d(CZ_4)^2$  operations  between nodes 1 and 2, and nodes 1 and 3, using one 3-qudit GHZ state instead of two entangled pairs of qudits, using a circuit similar to Figure~\ref{multitarget2}. Such a circuit for a distributed-$CSUM''_4$ is illustrated in Figure~\ref{multitarget2-dcsum4}.  On each node 2 and 3, we can then replace the local $CSUM_4$   with the corresponding  operations $(I \otimes H_4)~CSUM_4~CSUM_4~(I \otimes H_4^{\dagger})$ as we saw before to realize a $d(CZ_4)^2$ between the qudits on nodes 1 and 2 and the qudits on nodes 1 and 3. The circuit would need to use a 3-qudit GHZ state of the form $\sum_{k \in \{0,1,2,3\}} \frac{1}{2}\ket{kkk}$.
So one 3-qudit GHZ state (involving nodes 1, 2 and 3) and one entangled pair of qudits (involving nodes 2 and 3) suffices for a qudit based version of the 6-qubit GCZ operation.
This contrasts with the qubit based circuit, if distributed over three nodes, two qubits each, would require 12 entangled pairs  (4 between nodes 1 and 2, 4 between nodes 1 and 3, and 4 between nodes 2 and 3), if we use one pair for each (distributed) CZ between qubits on different nodes. 

{\bf Comparisons.} In summary, we compare three different implementations of the GCZ operation, based on the example we saw earlier with a 6-qubit GCZ operation and more generally, an $n$-qubit GCZ operation, as shown in Table~\ref{comp}, where $ep$ denotes an entangled pair of qubits, $ghz$ denotes a GHZ state (involving three or more qubits), $ep_d$ denotes an entangled pair of qudits of dimension $d$ and $ghz_d$ denotes a GHZ state of qudits of dimension $d$. An $n=kD$-qubit GCZ distributed over $D \geq 3$ nodes, $k$ qubits per node, will require $\frac{n(n-1)}{2} - D\frac{k(k-1)}{2} = \frac{n(n-k)}{2}$ dCZ gates, requiring that number of entangled pairs if each dCZ gate done separately. Using fan-out operations, we need $n-2k$ GHZ states (i.e., $k$ $D$-qubit GHZ states, $k$ $(D-1)$-qubit GHZ states,..., and $k$ $3$-qubit GHZ states), and $k$ entangled pairs. 
With multiple qubits per node, $O(n)$ GHZ states are still required since every qubit needs to interact with every other qubit for a $GMS/GCZ$ operation.
With qudit-based ``compression'' of the GCZ circuit, with qudit dimension $d=2^k$, we just have one qudit per node, and so require $(D-2)=(\frac{n}{k}-2)$ GHZ states, and one entangled pair.  For the case where qudit dimension is $d=2^m$ where $k=ml$, i.e. each qudit only  represents $m<k$ qubits, and so, $l$ qudits are required on each node; this means we need 
$(Dl-2l)=(\frac{n}{m}-2\frac{k}{m})$ (qudit-based) GHZ states and $l=\frac{k}{m}$ entangled pairs of qudits.
For example, with qudits of dimension $32=2^5$, perhaps a qudit compression ratio of 5 qubits per qudit  is possible, with a corresponding reduction in the number of GHZ states required (though now qudit based).

\begin{table*}[!htb]
 \caption{Comparison of entanglement resources for GCZ.}\label{comp}
\centering
 \begin{tabular}{|p{5cm} |c| c |p{6cm}|} 
 \hline
 Circuit Configuration & qubit, pairwise & qubit, using fan-out & compression with qudits of dimension $d$, using fan-out  \\ 
 \hline\hline
 6-qubit GCZ  distributed over 3 nodes, 2 qubits per node & 12 $ep$ & 2 $ghz$, 2 $ep$ & 1 $ghz_d$, 1 $ep_d$ \\ \hline
 $n$-qubit GCZ distributed over $D\geq 3$ nodes, with $kD = n$ and $k$ qubits per node  & $\frac{n(n-k)}{2}$ $ep$ & $(n-2k)$ $ghz$, $k$ $ep$ & $(\frac{n}{k}-2)$ $ghz_d$, 1 $ep_d$ (if $d=2^k$, 1 qudit per node) \;\; $(\frac{n}{m}-2\frac{k}{m})$ $ghz_d$, $\frac{k}{m}$ $ep_d$ (if $d=2^m$, where $k=ml$, for some integer $l$, $l$ qudits per node)   \\
 \hline
 \end{tabular}
\end{table*}

\section{Conclusion and Future Work}
\label{sec4}
We have seen that  GHZ states can be used for distributed fan-out operations, and so, global gates comprising sequences of such fan-out operations can be distributed. We considered qudit compression of qubit circuits for global (GCZ) gates, so that distributed GCZ operations can be realized with distributed fan-out operations; we compare this approach, where possible, with naive realizations of such global gates using long sequences of dCNOT operations, arguing that using distributed fan-out qudit operations can be more efficient provided single-shot multipartite entanglement can be  efficient. 

 Future work on compiler design for distributed quantum circuits, and quantum data centres (QDCs), include:
\begin{itemize}
\item  distributed quantum computing circuits could be compiled into circuits using not just bipartite entangledd pairs but also  multipartite entanglement (e.g., for fan-out operations involving three or more qubits), if forming such multipartite entangled states like GHZ states in a single shot can be nearly as efficient as forming bipartite entangled pairs, then distributed fan-out operations can be very efficient - gate reordering~\cite{Mengoni:2025lsk} (e.g., exploiting commuting operations)  can perhaps be adapted to reorder dCNOT gates into sequences which can be replaced by equivalent distributed fan-out operations (i.e.,  one shot fan-out execution versus multiple  dCNOTs);

\item  apart from generators of bipartite entangled pairs, QDCs could also have efficient one shot  generators of multipartite entangled states (e.g., GHZ states or W states); e.g., switches connecting different nodes on a QDC such as those mentioned in \cite{10.1145/3649329.3655908,10.1145/3695053.3731046} might  be extended  to generate distributed GHZ states, requiring scheduling for both EPR pairs and GHZ states generation;

\item the number of communication qubits (or qudits) per node  is a design consideration, since if nodes have more communication qubits (to hold entanglement resources), then more distributed operations can be executed concurrently; but an approach is  in ~\cite{10.1116/5.0241972} which considered shared pool of entanglement resources, to cope with different needs of communication on different nodes depending on the quantum circuit and qubit allocation;



\item  qudit based compression of qubit circuits can be utilised either for efficient quantum information transfer (if qudit teleportation  can be made almost as efficient as qubit teleportation), or if entangled states of qudits can be formed efficiently,   qudits can be used to compress qubit circuits enabling reduced depth execution and possibly more efficient distributed quantum operations; 

\item different quantum nodes/servers interconnected might use different quantum technologies with different strengths  - compiling for such heterogeneous nodes    can be challenging, as already mentioned by others (e.g., ~\cite{Bandic2025,9334411}); and

\item apart from EPR pairs and GHZ states,  $W$ states   (e.g., $\frac{1}{\sqrt{3}} (\ket{001}+\ket{010}+\ket{100})$) might be a useful ``basic'' resource in QDCs, which can be used for distributed coordination  such as leader election and  advantages for   multiqubit CHSH games~\cite{Bandic2025}, and more robust than GHZ states under qubit loss. Qudit generalizations exist~\cite{10.1145-3594671.3594687}.
\end{itemize}



\bibliographystyle{plain}
\bibliography{refs}
 
\end{document}